# Optically Cooling Cesium Lead Tribromide Nanocrystals

*Benjamin J. Roman, Noel Mireles Villegas, Kylie Lytle, Matthew T. Sheldon\**

Benjamin J. Roman, Noel Mireles Villegas, Kylie Lytle, Dr. Matthew T. Sheldon
Department of Chemistry, Texas A&M University, College Station, TX, 77843, USA
E-mail: sheldonm@tamu.edu

Dr. Matthew T. Sheldon
Department of Materials Science & Engineering, Texas A&M University, College Station, TX, 77834, USA

**Abstract:**

One photon up-conversion photoluminescence is an optical phenomenon whereby the thermal energy of a fluorescent material increases the energy of an emitted photon compared with the energy of the photon that was absorbed. When this occurs with near unity efficiency, the emitting material undergoes a net decrease in temperature—so called optical cooling. Because the up-conversion mechanism is thermally activated, the yield of up-converted photoluminescence is also a reporter of the temperature of the emitter. Taking advantage of this optical signature, cesium lead trihalide nanocrystals are shown to cool during the up-conversion of 532 nm CW laser excitation. Raman thermometric analysis of a substrate the nanocrystals were deposited on further verifies the decrease in the local environmental temperature by as much as 25 °C during optical pumping. This is the first demonstration of optical cooling driven by colloidal semiconductor nanocrystal up-conversion.

**Main Text:**

One photon up-conversion, also known as anti-Stokes photoluminescence (ASPL), is a phonon-mediated up-conversion mechanism where thermal energy from the emitting material combines with a low energy absorbed photon to produce higher energy optical emission. If this process occurs with an external quantum efficiency (EQE) near unity, more thermal energy is removed via ASPL than is added by thermalization due to nonradiative recombination, and the material will experience a net decrease in temperature. This mechanism of optically-pumped cooling, also called laser cooling,[1] has been demonstrated for a number of materials over the decades since it was first proposed.[2] Fluorescent dyes, for example, have been demonstrated to cool by several degrees during ASPL.[3] Most commonly, rare earth metal doped glasses are used for studying and demonstrating optical cooling, reaching temperatures as low as 91 K,[4] close to the thermodynamic limit of that material system.[5]

In comparison, semiconductors are theoretically predicted to be able to reach temperatures below 10 K via ASPL optical cooling;[1,6] despite this, optical cooling of bulk semiconductors has yet to be demonstrated. For net cooling to occur, the thermal energy scavenged by emitted photons—a few tens to a few hundreds of meV worth of energy for each emission event—has to be greater than the thermal energy generated by non-radiative losses, with each non-radiative recombination contributing a full band gap worth of heat energy. As such, the EQE of the semiconductor must approach unity for optical cooling to occur. Despite having internal quantum efficiencies (IQE) exceeding 99%, high quality bulk semiconductors still have to contend with insufficient photon extraction efficiencies due to total internal reflection and parasitic absorption from their surface passivation, losses that are largely an intrinsic feature of the macroscopic semiconductor geometry.[7]

In recent years, the first instances of measurable laser cooling of semiconductors have been reported by Xiong and coworkers. In their experiments, chemical vapor deposition was used to fabricate semiconductor morphologies that are subwavelength in size in at least one dimension, in order to maximize the optical extraction efficiency of the ASPL, i.e. to promote high EQE *via* better light management.[8] Until our study here, their reports remained the only demonstrations of net optical cooling of a semiconductor, largely due to the difficulty of consistently and reproducibly fabricating materials with sufficiently high EQE.

High quality colloidal semiconducting nanocrystals appear to ubiquitously show ASPL [9] and can be synthesized in subwavelength sizes with EQE above the requisite threshold for net cooling.[10] Notably, recent works have identified all-inorganic cesium lead trihalide perovskite nanocrystals as a material with potential applications for optical cooling due to their near-unity EQE after appropriate surface treatment, making them ideal candidates for demonstrating optical cooling.[11–14]

Here, we report for the first time the optically driven cooling of colloidally prepared semiconducting nanocrystals—specifically $CsPbBr_3$ nanoparticles. We use the known Arrhenius behavior of the ASPL yield to estimate the change in temperature of the nanocrystals during below gap excitation. The rate of cooling, as well as the final temperature reached, are shown to be dependent on the excitation laser fluence. Additionally, we verify the temperature change of the environment around the nanocrystals by monitoring the anti-Stokes to Stokes Raman scattering ratio of a silicon substrate on which $CsPbBr_3$ nanoparticles were deposited. This analytical thermometry technique measures the temperature-dependent phonon mode population of silicon to report the local temperature of the substrate.

The necessary EQE threshold for cooling can be calculated by considering the amount of thermal energy emitted with each photon, as defined by the anti-Stokes shift between the excitation wavelength and the emission wavelength.[1] In our study, 532 nm excitation of $CsPbBr_3$ produces PL centered between 513 to 517 nm depending on the sample (**Figure S1**). This corresponds to an anti-Stokes shift of approximately 70 to 90 meV. Our work and others' have established this as a one photon up-conversion process whereby absorption of multiple phonons converts below-bandgap absorption into band edge emission.[11,12,14,15] Thus, for every photon emitted, approximately 70 to 90 meV of thermal energy is removed from the semiconductor. In contrast, every instance of non-radiative recombination adds 2.4 eV of thermal energy into this system. For ASPL to remove more thermal energy than the thermal energy added through non-radiative recombination, ASPL emission into free space must occur with greater than ~97% efficiency.

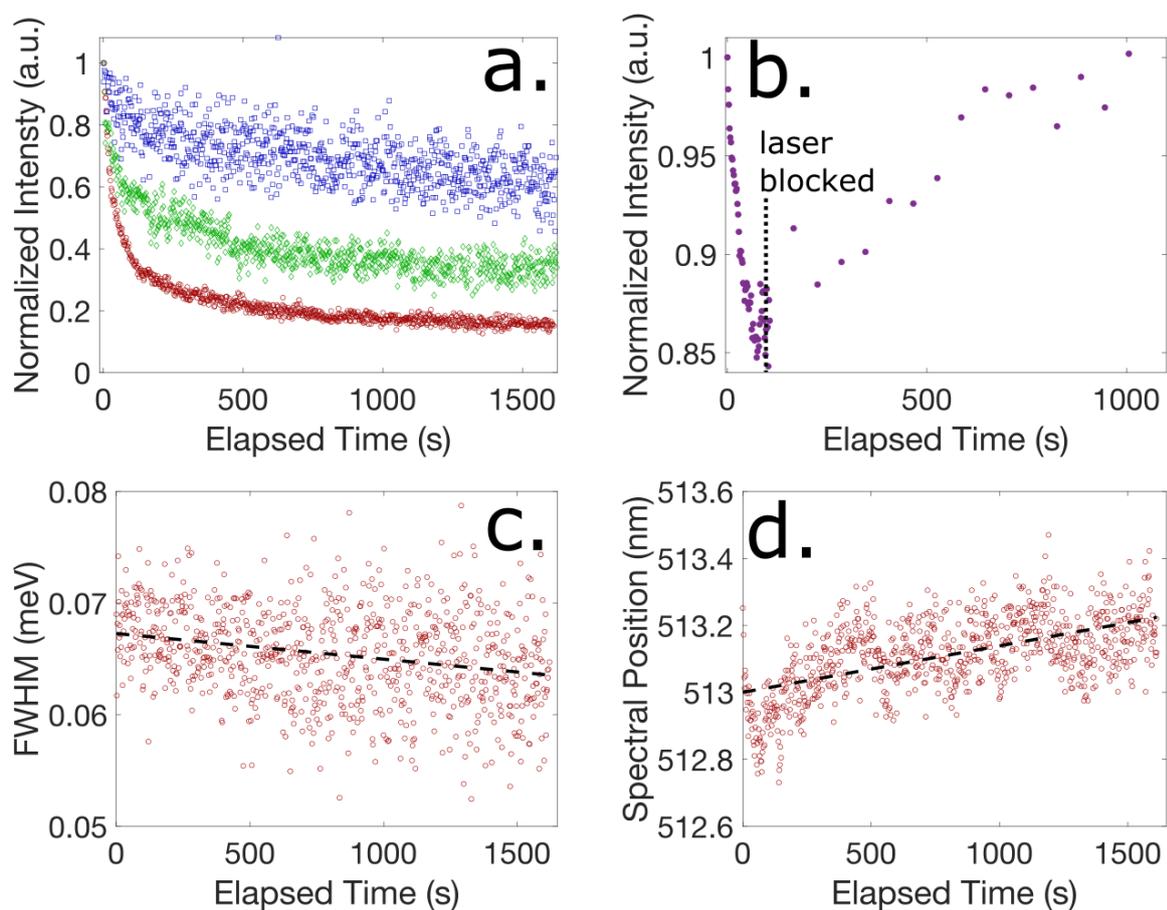

**Figure 1.** CsPbBr$_3$ anti-Stokes photoluminescence spectral changes during below-gap (532 nm) excitation. (a) Excitation fluence dependent decrease in ASPL intensity over time. Blue squares, green diamonds, and red circles correspond to excitation fluences of 300, 1500, and 3000 W/cm$^2$ respectively. (b) CsPbBr$_3$ pumped below-gap (532 nm) at a fluence for 15 W/cm$^2$ for 100 seconds. The laser was then blocked, as indicated by the dotted line. The laser was unblocked periodically to check the ASPL intensity. (c) ASPL decrease in full-width at half-max with an excitation fluence of 3000 W/cm$^2$. (d) Red shift of ASPL spectral position with an excitation fluence of 3000 W/cm$^2$. In (c) and (d) the dashed line is a linear fit to the data. These data sets are from the same experiment as the red circles in (a).

CsPbBr$_3$ nanoparticles were synthesized following the hot-injection method established by Protesescu and coworkers.[16] They were then treated with NH$_4$SCN to increase their EQE to approximately unity. In this reaction, SCN$^-$ removes the excess lead atoms on the nanocrystal surface that are understood to be the predominant source of mid-gap states.[17] After treatment with NH$_4$SCN, our sample was confirmed to have 98.5 ± 4% EQE measured against a fluorescein standard.[18] To demonstrate cooling, these nanoparticles were mixed into a 5% solution of polystyrene in toluene, drop-cast on a quartz slide, and placed under vacuum. Both the polystyrene encapsulation and the vacuum were used to thermally isolate the nanoparticles, reducing their thermal load to maximize the observed cooling. The sample was then excited using 532 nm CW laser excitation, and the resultant ASPL spectrum was collected continuously or at regular intervals during the course of an experiment.

During below-gap excitation, the collected ASPL spectra changed over time, decreasing in intensity as well as often undergoing a decrease in its full-width at half-max (FWHM), and a red-shift of the spectral position of the photoluminescence (PL) peak (**Figure 1**). Figure 1a highlights the decrease in ASPL intensity as a function of the excitation fluence, with greater light intensity leading to a faster rate of decrease. Given enough time, the ASPL stabilized at a saturation point, where it remained steady so long as the excitation fluence was not increased or decreased (**Figure S2**). Most tellingly, this change in the ASPL intensity was entirely reversible, either through blocking the laser (Figure 1b) or through decreasing the excitation fluence once the ASPL decay reached its saturation point (Figure S2). The reversibility of the change in ASPL intensity is an important characteristic that differentiates the spectral changes observed here from non-reversable photodegradation that has been observed when CsPbBr$_3$ nanoparticles are exposed to trace water, O$_2$, or other environmental factors that degrade their electronic structure.[19]

ASPL is a thermally activated process. As such, ASPL yield increases with increasing temperature (**Figure 2** and S1), and conversely decreases with decreasing temperature. Thus, we hypothesize that the observed decrease in the intensity of the ASPL spectra is due to the thermal deactivation of the nanoparticles brought about by a net decrease in temperature. Additional verification that this decrease in ASPL signal corresponds to a decrease in temperature can be seen in the FWHM of the ASPL spectra (Figure 1c). A decrease in the FWHM of the ASPL peak over the course of the measurement is consistent with a decrease in the thermal activation of carriers participating in optical recombination across the semiconductor band gap. Similarly, a red-shift of the ASPL peak position (Figure 1d), also indicates thermal de-activation of the energy distribution of carriers recombining across the band gap as temperature is decreased.[20,21]

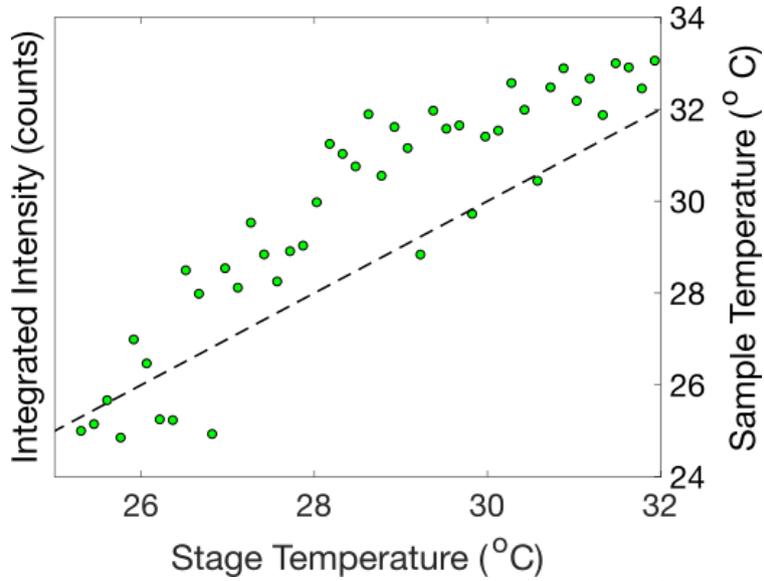

**Figure 2.** The temperature reported by the internal thermocouple of a heating stage plotted versus the temperature (right axis) estimated using the change in CsPbBr$_3$ ASPL intensity (left axis) according to equation 1. The dashed line corresponds to the temperature reported by the thermocouple.

During ASPL, each up-converted emission event can be thought of as a cooling cycle that removes thermal energy by depopulating the phonon modes of the nanoparticle. The depopulation of the phonon modes in turn reduces the ASPL yield and the amount of thermal energy that is removed per unit time, even under a constant laser fluence. This photo-induced thermal deactivation is manifest in the fluence dependence of the rate of the ASPL signal decay we observe: Higher laser fluences remove a greater amount of thermal energy per unit time, causing a faster decay in the ASPL intensity (Figure 1a). After significant thermal deactivation, the ASPL yield becomes too low to overcome the thermal energy entering the nanocrystals from their environment, and the ASPL intensity approaches a constant value (Figure S2). This constant ASPL intensity corresponds to a steady state when thermal energy flowing into the nanocrystals from the environment is equal to the thermal energy removed per unit time via ASPL. Critically, the observed reversibility of this decay signal is due to this thermal dependence of the ASPL mechanism. When the ASPL is no longer being pumped (i.e. the laser is blocked) or the rate of heat removal via ASPL is decreased such that it no longer counterbalances the thermal energy from the nanoparticles' environment (i.e. the laser fluence is decreased), the nanoparticles increase in temperature and the ASPL intensity increases as a consequence. In the former case, when the laser is blocked, the nanoparticles will warm back up to room temperature, as demonstrated in Figure 1b. In the latter case, when the laser fluence is decreased, the ASPL intensity increases until the nanoparticles once more reach a steady state, and the ASPL intensity remains constant (Figure S2) with the nanocrystals at a temperature once again dictated by the rate balance of heat removal due to ASPL and heat provided from their environment.

The temperature dependence of the ASPL intensity from CsPbBr$_3$ nanocrystals is known to follow an Arrhenius relation, with the natural log of the intensity directly proportional to 1/T.[11] Thus, the change in temperature required to bring about a given change in ASPL intensity can be estimated using equation 1, which is simply the ratio of two Arrhenius equations solved for some unknown temperature:

$$T = [\frac{k_B}{E_a} \ln\left(\frac{I_0}{I}\right) + \frac{1}{T_0}]^{-1} \qquad (1)$$

In this, $I_0$ is the integrated ASPL intensity at some known temperature, $T_0$; $I$ is the integrated ASPL intensity at some new temperature, $T$; $k_B$ is Boltzmann's constant, and $E_a$ is the energy of activation of the one photon up-conversion process. We have previously estimated the ASPL energy of activation for CsPbBr$_3$ by heating the nanocrystals in a thermal stage.[11] Alternatively, for this study we cooled a thin film sample of nanocrystals using an ice water bath in order to ensure that the estimated energy of activation was comparable for the temperature range over which optical cooling was observed. The energy of activation was estimated to be ~140 meV (**Figure S3**), in line with previously published measurements.[11,21]

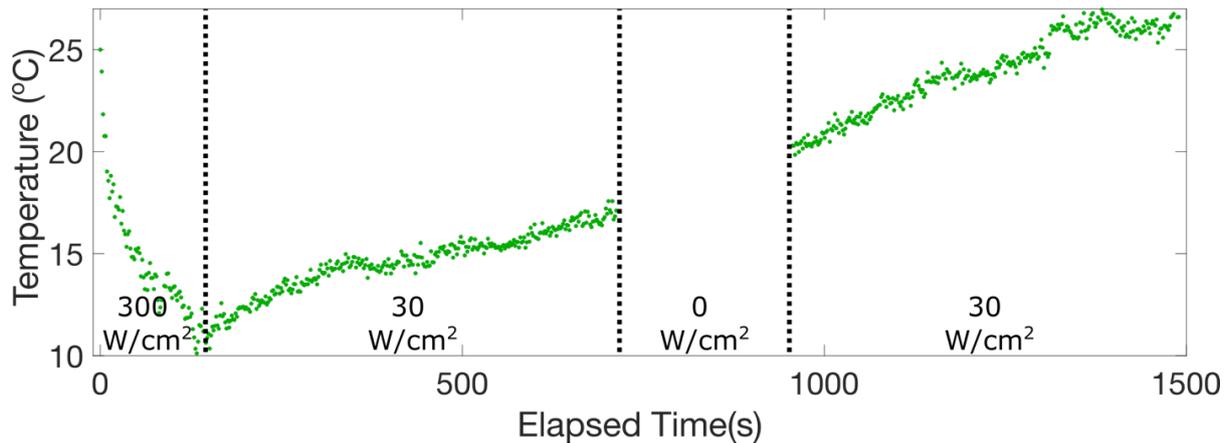

**Figure 3.** Monitoring the temperature of CsPbBr$_3$ nanocrystals during below-gap excitation (532 nm). The nanoparticles were cooled at an excitation fluence of 300 W/cm$^2$, followed by a decreased fluence of 30 W/cm$^2$, the laser being blocked, and then 30 W/cm$^2$ again.

As a check of equation 1, a sample of CsPbBr$_3$ nanocrystals was heated on a temperature-controlled heating stage while simultaneously collecting ASPL spectra. The CsPbBr$_3$ nanocrystals were not treated with NH$_4$SCN to ensure that the nanoparticles' EQE was below the threshold for optical cooling. The temperature of the nanoparticles, estimated using equation 1, was within a few degrees of the temperature reported by the thermal stage's internal thermocouple, a difference that may be due to local variations in the sample temperature (Figure 2). This confirms that equation 1 can accurately estimate the temperature of the nanoparticles, at least over the temperature range for which the ASPL energy of activation has been calibrated (10 ºC to 25 ºC).

For the purposes of estimating the temperature of the nanocrystals during below-gap excitation, the first recorded spectrum for a given spot on the sample is assumed to be approximately room temperature. A decrease—or increase—in ASPL intensity can then be correlated to a change in temperature according to equation 1. **Figure 3** demonstrates a typical cooling experiment. The sample was pumped below-gap with a fluence of 300 W/cm$^2$ such that thermal energy was removed faster than it was replaced by the environment surrounding the nanocrystals in the optical spot. The temperature began to drop exponentially, reaching an estimated temperature of approximately 10 °C after 2 minutes. This temperature is within the temperature range over which we calibrated the ASPL energy of activation. The excitation fluence was then decreased to 30 W/cm$^2$, below the fluence threshold necessary to overcome the heat flux from the environment into the nanocrystals at this particular spot on the sample (**Figure S4**). Therefore, the temperature of the nanoparticles began to increase. The laser was additionally blocked for 5 minutes to demonstrate the continued reversal of the decay in ASPL intensity, even in the dark. After approximately 20 minutes, the nanoparticles returned to room temperature. Figures 1a and 1b are similarly reproduced as nanoparticle temperature versus time in **Figure S5** and **Figure S6**, respectively.

The red shift, decrease in FWHM, and reversible decrease in intensity of ASPL spectra all strongly suggest that the nanoparticles are experiencing a net decrease in temperature. For additional verification of this hypothesis, we also sought evidence that the nanoparticles were cooling their local environment during ASPL. For this, we took advantage of the well know temperature dependence of Raman scattering from a Si substrate the nanoparticles were deposited on. Silicon has a strong Raman scattering peak at 520 cm$^{-1}$. The ratio of the anti-Stokes and Stokes Raman peaks corresponding to this vibrational mode can be used to determine the absolute temperature of the silicon, as described in equation 2, where $I_{AS}$ and $I_S$ are respectively the silicon anti-Stokes and Stokes scattering intensities, $v_l$ and $v_v$ are respectively the frequencies of the laser and the silicon vibrational mode, $h$ is Planck's constant, $k_B$ is Boltzmann's constant, and $T$ is the temperature of the silicon.[22]

$$\frac{I_{AS}}{I_S} = \left(\frac{v_l+v_v}{v_l-v_v}\right)^3 e^{\frac{-hv_v}{k_B T}} \qquad (2)$$

As the temperature of the silicon decreases, the phonon mode corresponding to $v_v$ is depopulated, and the anti-Stokes scattering peak decreases in intensity relative to the Stokes scattering peak. It is important to note that equation 2 does not require prior calibration, as it directly correlates the temperature dependence of the $v_v$ phonon mode population with the ratio of the accompanying Stokes and anti-Stokes scattering cross sections.

To confirm that the nanoparticles were indeed decreasing the local temperature, $NH_4SCN$ treated $CsPbBr_3$ nanoparticles were dropcast onto a cleaned silicon wafer and placed under vacuum. The nanoparticles were excited at a focal plane just above the substrate surface that maximized ASPL intensity, so as to maximize the thermal energy removed per unit time. The silicon Raman signal could be resolved simultaneously to the ASPL intensity (**Figure S7** and **Figure S8**). The temperature estimated by the change in ASPL according to equation 1 is plotted in **Figure 4** alongside the temperature determined by the silicon anti-Stokes to Stokes ratio according to equation 2. Note that after 160 seconds, the anti-Stokes silicon peak was too low in intensity to be resolved against the noise. After ~800 seconds, the $CsPbBr_3$ ASPL intensity began to increase again, possibly indicating that the nanoparticles were being damaged under the high fluence necessary to resolve the silicon Raman peaks ($10^5$ W/cm$^2$) and were beginning to heat. At the end of the measurement, the focal plane was optimized so that the silicon Raman signal could once more be resolved with a good signal-to-noise ratio in order to measure the final temperature of the silicon substrate.

The temperature reported by the silicon Raman scattering is remarkably close to the temperature estimated using the $CsPbBr_3$ ASPL, with the silicon just a few degrees higher than the nanoparticles themselves. The final temperature of the silicon was determined to be -1.7 °C, as compared to the final temperature of the nanoparticles, estimated to be -5 °C. Note that the change in the temperature of the silicon is certainly a locally induced change in the vicinity of the optical spot. Further study and optimization will be necessary to use CsPbBr$_3$ nanoparticles to induce a net decrease in temperature for a bulk thermal load.

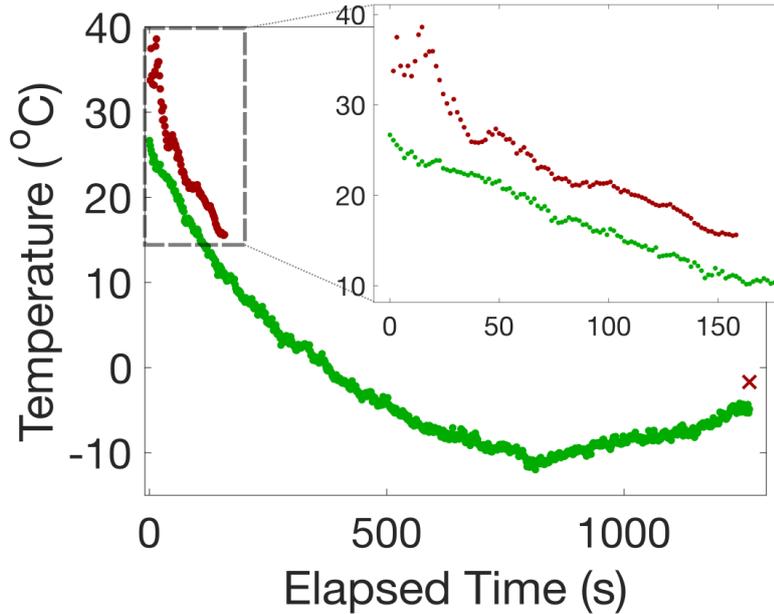

**Figure 4.** Temperature of CsPbBr$_3$ nanoparticles during below gap excitation, estimated using the CsPbBr$_3$ ASPL intensity (green) as well as the anti-Stokes to Stokes Raman scattering ratio of Si (red). At the end of the measurement, the focus was adjusted to maximize the silicon Raman scattering collection efficiency. The final temperature of the silicon substrate is reported as the red x.

In conclusion, we analyzed both the temperature dependent yield of ASPL and the anti-Stokes to Stokes Raman scattering ratio of a silicon substrate to confirm that below gap excitation of CsPbBr$_3$ nanoparticles can induce a net decrease in temperature. A remarkable aspect of these experiments is the reproducibility and consistency with which CsPbBr$_3$ nanoparticles can be optically cooled. While the rate and magnitude of the cooling are dependent on the local environment and thermal insulation of each spot analyzed in our experiments, nearly every measurement showed an exponential and reversible decrease in nanoparticle temperature. It is currently unclear whether the ease with which CsPbBr$_3$ nanoparticles exhibit optical cooling is a result of their unique photophysical characteristics, or whether cooling should be expected as a general feature of ASPL with near-unity EQE. Certainly, the observed cooling is comparable in magnitude and timescale to that reported by Xiong and co-workers.[8] The comparison between studies suggests that the optical cooling may, in fact, be a feature of the high EQE, and the optical extraction efficiency afforded by the sub-wavelength geometry. As such, colloidal semiconductor nanoparticles with high quantum yield may likely provide an ideal platform for the study and application of optical cooling moving forward.

**Experimental Section**

*Synthesis of CsPbBr$_3$:* $Cs_2CO_3$ (0.200 g), OA (0.624 mL), and ODE (10 mL) were added to a 25-mL 3-neck round bottomed flask and heated for 1 hour at 120°C under vacuum to dry. After 1 hour, the flask was put under argon and heated to 150°C until all the $Cs_2CO_3$ had reacted. $PbBr_2$ (0.069 g) and ODE (5 mL) were added to a 25-mL 3-neck round bottomed flask and heated under vacuum to 120 °C for 1 hour. The solution was then placed under argon, and dried OAm (0.5 mL) and dried OA (0.5 mL) were injected to solubilize the $PbBr_2$. The temperature increased to 180°C, and the Cs-oleate (0.4 mL) was swiftly injected. After 1 minute, the solution was cooled with an ice bath. The final solution was centrifuged at 3000 g-forces for 5 minutes and the supernatant was discarded. The precipitate was dispersed in hexane.

*$NH_4SCN$ Treatment of CsPbBr$_3$ Nanoparticles:* $NH_4SCN$ was added to a vial of $CsPbBr_3$ suspended in hexane and vigorously stirred for between 20 and 30 minutes. The resulting cloudy solution was centrifuged at 3000 g-forces for 5 minutes, and the supernatant was decanted and analyzed.

*Determination of sample EQE:* Sample EQE was determined versus a standard solution of fluorescein in 0.1 M NaOH. The absorbance of a dilute solution was collected on an Ocean Optics Flame-S-UV-Vis spectrometer with an Ocean Optics DH-200-Bal deuterium and halogen lamp light source. Photoluminescence was collected on the same Ocean Optics Flame-S-UV-Vis spectrometer, using a Fianium WhiteLase supercontinuum laser ported through an LLTF high contrast filter as the excitation source.

*Calibration of ASPL Energy of Activation:* 100 μL of $NH_4SCN$ treated $CsPbBr_3$ was diluted with 1 mL of hexane. Several drops were sandwiched between two glass slides and sealed inside of a plastic bag in an argon filled glove box. The sample was clamped in place in an ice bath at a 45 degree angle to the 532 nm CW laser excitation source. Two Edmund Optics OD4 532 nm notch filters were used to block scattered light from the laser. A Stanford SR830 lock-in connected to a Thor Labs DET100A2 detector was used to collect the sample ASPL. The temperature of the sample was measured with a Digi-sense TC9000 advanced temperature controller and a temperature probe in the ice bath. The ice bath was stirred with a magnetic stir bar to ensure uniformity in temperature. Once the ice in the ice bath had melted, the ASPL was periodically measured as the water bath and sample warmed up to room temperature. The laser was blocked in between measurements.

*Measurement of below-gap excitation photoluminescence (ASPL):* ASPL was measured using a 532 nm Nd:YAG CW laser ported through a WITec alpha 300 RA confocal microscope, focused on the sample using a long working distance 20x objective with a 0.35 numerical aperture. Between 5 and 100 μL of $NH_4SCN$ treated $CsPbBr_3$ was added to 1 mL of a 5% by w.t. solution of polystyrene in toluene. One drop was placed on a quartz slide and allowed to dry. The sample was then placed in a Linkam Instruments TS1500 thermal stage attached to the WITec alpha 300 RA confocal microscope. Measurements were taken at a vacuum pressure 0.010 mBar.

*Measurement of silicon Raman scattering*: 10 μL of $NH_4SCN$ treated $CsPbBr_3$ was diluted with 500 μL of hexane and dropcast on a clean silicon substrate. The sample was then analyzed as described above for the measurement of below-gap excitation photoluminescence, except with a 20x near working distance objective with a numerical aperture of 0.4.

**Acknowledgements**

This work is funded by the Gordon and Betty Moore Foundation through Grant GBMF6882. M.S. also acknowledges support from the Welch Foundation (A-1886) and the Air Force Office of Scientific Research under award number FA9550-16-1-0154.

# Supporting Information

**Optically Cooling Cesium Lead Tribromide Nanocrystals**

*Benjamin J. Roman, Noel Mireles Villegas, Kylie Lytle, Matthew T. Sheldon**

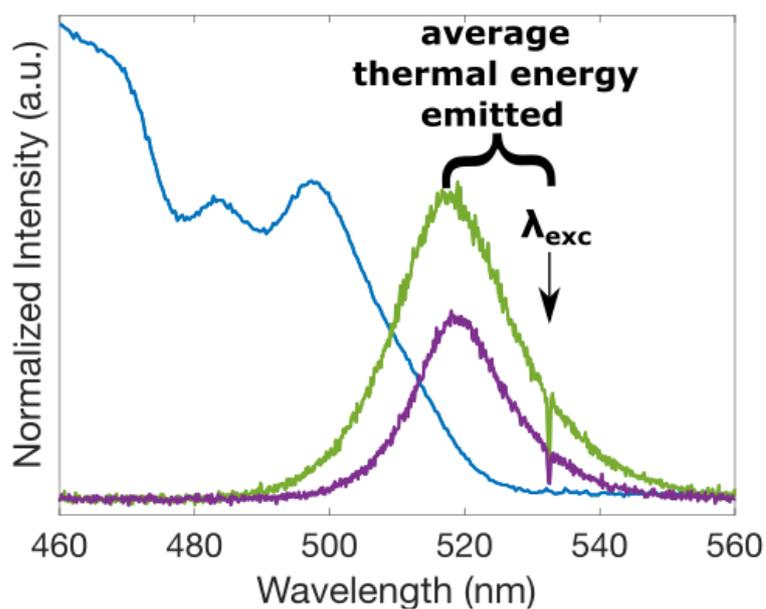

**Figure S1.** CsPbBr$_3$ absorbance (blue) and anti-Stokes photoluminescence when the nanocrystals are at 25 °C (purple) and 65 °C (green).

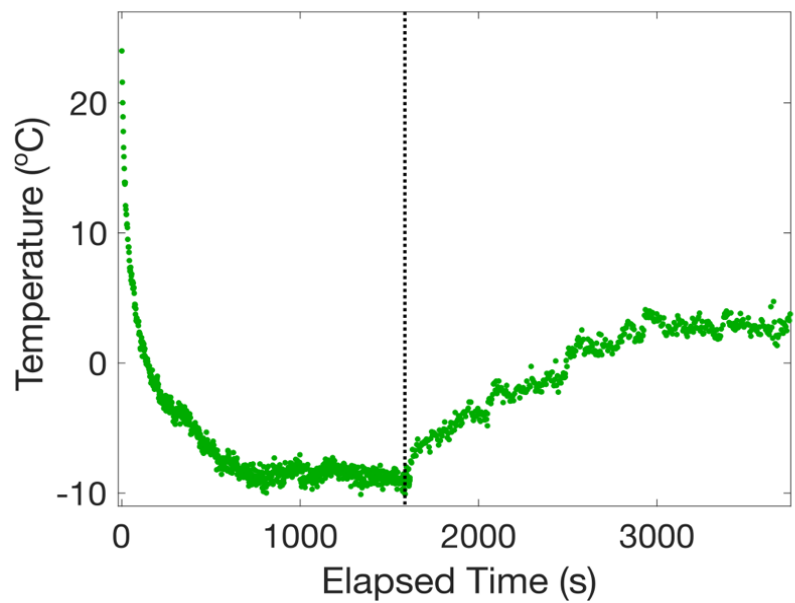

**Figure S2**. NH₄SCN treated CsPbBr₃ excited below gap (532 nm) at a fluence of 4600 W/cm² until the temperature reached a steady state. The laser fluence was then decreased to 2300 W/cm² until the temperature once again reached a steady state. The dotted line indicated when the laser fluence was decreased. The temperature was estimated using equation 1.

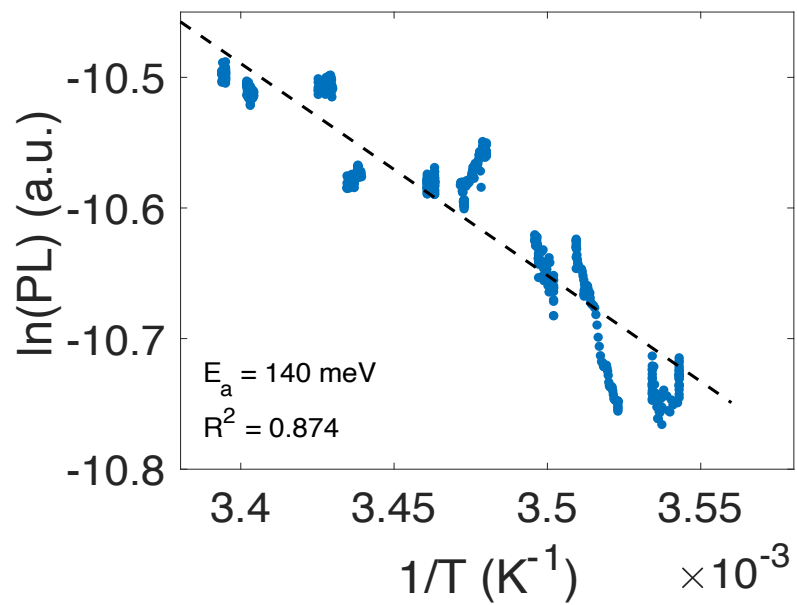

**Figure S3.** An Arrhenius plot of the temperature dependence of NH₄SCN treated CsPbBr₃ ASPL over the temperature range 10–24 °C. The estimated energy of activation was 140 meV.

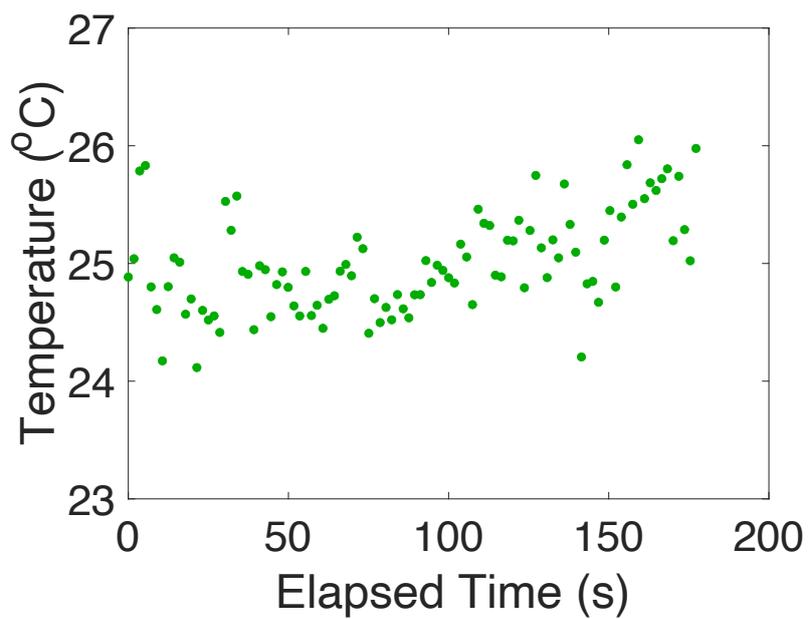

**Figure S4.** NH$_4$SCN treated CsPbBr$_3$ nanocrystals, pumped below gap (532 nm) at a fluence of 30 W/cm$^2$. The steady ASPL intensity, and thus the temperature estimated using equation 1, indicates that this fluence is too low to overcome heat flux from the nanoparticles' environment. This is the same sample analyzed in Figure 3.

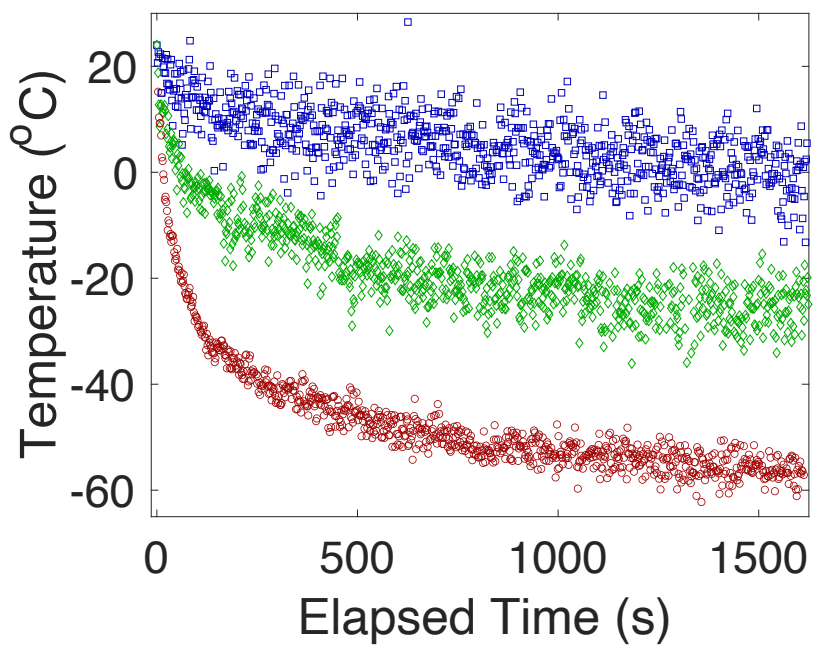

**Figure S5.** CsPbBr$_3$ anti-Stokes photoluminescence spectral changes during below-gap (532 nm) excitation. Excitation fluence dependent decrease in temperature over time. Blue squares, green diamonds, and red circles correspond to excitation fluences of 300, 1500, and 3000 W/cm$^2$ respectively. This data set is the same as in Figure 1a, using equation 1 to estimate temperature. Note that the temperatures estimated here are only accurate so long as the ASPL energy of activation holds constant over this temperature range.

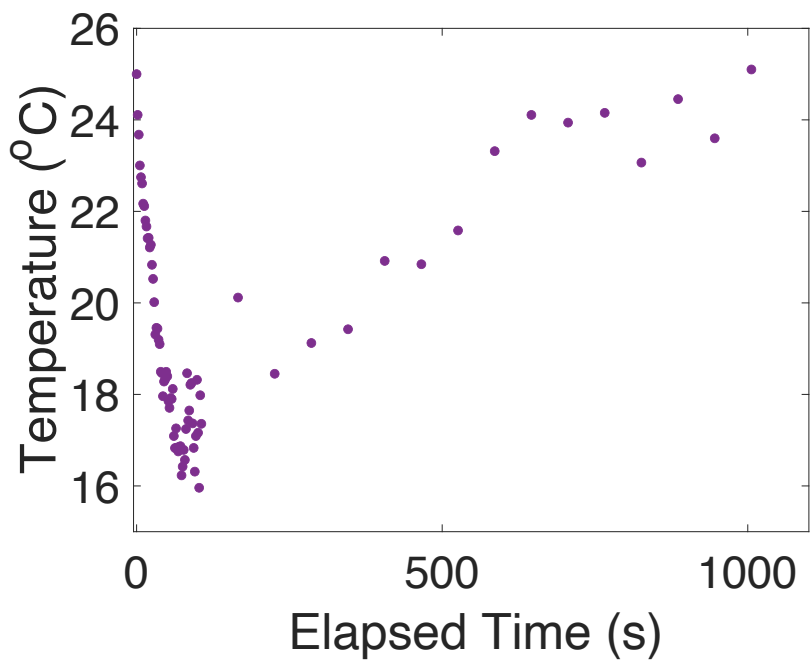

**Figure S6.** CsPbBr$_3$ pumped below-gap (532 nm) at a fluence for 15 W/cm$^2$ for 100 seconds. The laser was then blocked. The laser was unblocked periodically to check the ASPL intensity. This is the same data set as in Figure 1b, using equation 1 to estimate temperature.

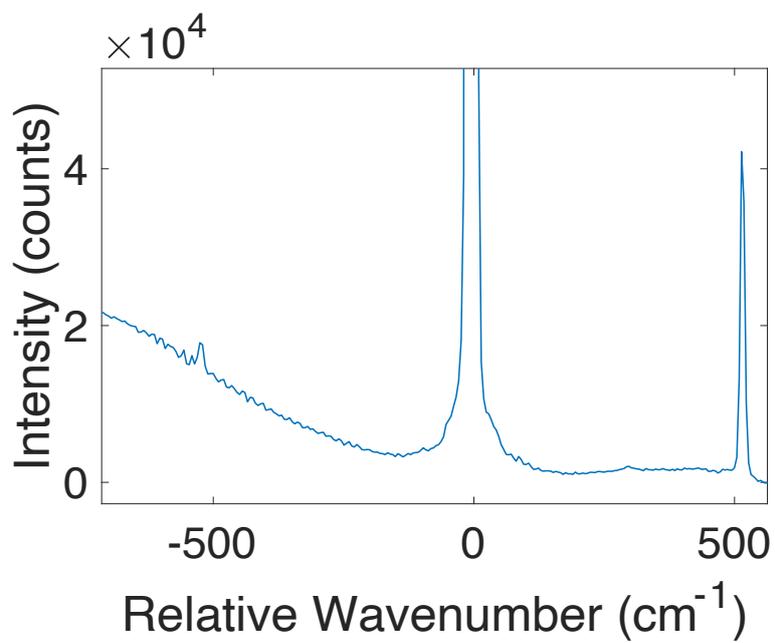

**Figure S7.** Silicon Raman scattering resolved simultaneously to CsPbBr$_3$ ASPL.

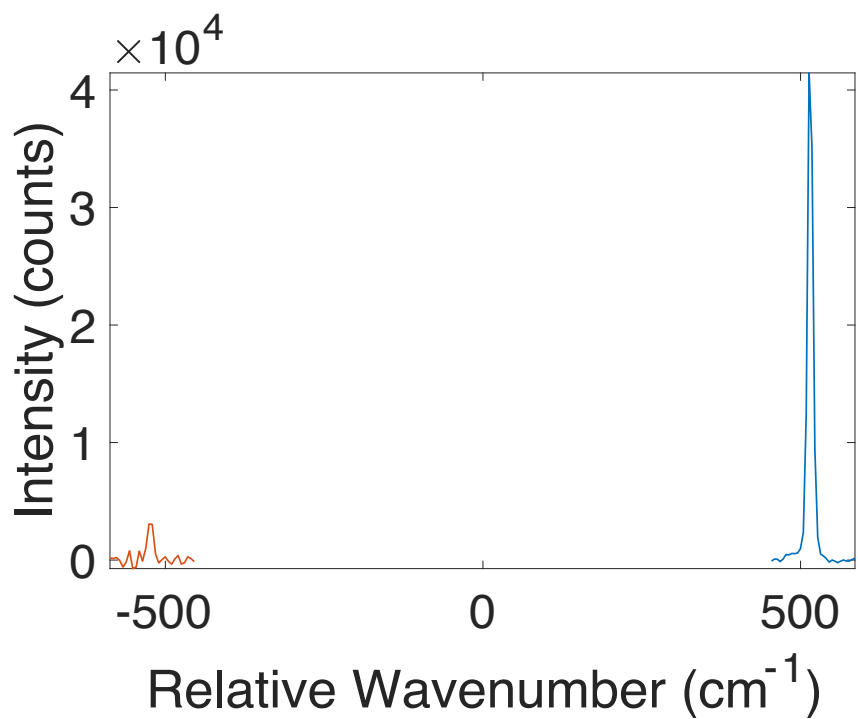

**Figure S8.** Silicon Raman scattering signal after subtracting the background. Orange is the anti-Stokes scattering signal, and blue is the Stokes scattering signal. This is the same data set as in Figure S7.